\documentclass[pss]{wiley2sp}
\usepackage[latin1]{inputenc}
\usepackage{amsmath, mathrsfs, amssymb, amscd}

\newcommand{\eps}{\varepsilon_1}  
\newcommand{\sig}{\sigma_1}       
\newcommand{\Cg}{\triangle}      
\newcommand{\rhop}{r_\omega}     
\newcommand{\WW}{U(\rhop)}       
\newcommand{\quant}{\hbar\omega}    
\newcommand{\dc}{\sigma_{\rm dc}}     

\begin{document}

%
%

\title{Influence of electronic correlations on the frequency-dependent
hopping transport in Si:P}

\titlerunning{ac hopping transport in Si:P}

\author{%
  Elvira Ritz and
  Martin Dressel}

\authorrunning{Elvira Ritz and Martin Dressel}

\mail{e-mail
  \textsf{ritz@pi1.physik.uni-stuttgart.de}, Phone
  +49-(0)711-68564893, Fax +49-(0)711-68564886}

\institute{%
  \,1. Physikalisches Institut, Universität Stuttgart, Pfaffenwaldring 57, 70550 Stuttgart, Germany}

\received{6 August 2007, revised 17 September 2007, accepted 18
September 2007} \published{7 February 2008}


\abstract{%
%
%
\abstcol{%
  At low energy scales charge transport in the insulating Si:P is dominated
  by activated hopping between the localized donor electron states. Thus, theoretical
  models for a disordered system with electron-electron interaction are appropriate
  to interpret the electric conductivity spectra. With a newly developed technique
  we have measured the complex broadband microwave conductivity of Si:P from 100 MHz
  to 5 GHz in a broad range of phosphorus concentration $n/n_c$ from 0.56
  to 0.95 relative to the critical value $n_c=3.5\times 10^{18}$ cm$^{-3}$
  corresponding to the metal-insulator transition driven by doping.}{At our base temperature
  of \,$T$\,=\,1.1 K the samples are in the zero-phonon
  regime where they show a super-linear frequency dependence of the conductivity indicating the influence of the
Coulomb gap in
  the density of the impurity states. At higher doping $n\rightarrow n_c$, an abrupt drop in the conductivity power law $\sig(\omega)\sim\omega^\alpha$
  is observed. The
  dielectric function $\eps$ increases upon doping fol-lowing a power law in ($1-n/n_c$). Dynamic response at elevated temperatures has also
  been investigated.
}}

\maketitle

\section{Introduction}
In a material of such an industrial importance like Si:P there are
still questions open about the low-energy excitations from the
ground state. In particular, the influence of electron-electron
interactions on the hopping transport and the critical behavior at
the metal-insulator transition (MIT) have attracted much attention
since decades. Though theoretical predictions on the
$T$$\rightarrow$0 frequency-dependent response of an interacting
disordered system have already existed since many years \cite{ES},
experimental data still remain scarce and results obtained by
different groups and in different parameter ranges lack
consistency. Especially in the microwave range, from tens of MHz
till tens of GHz, with wavelengths from a few millimeters to a few
meters, no better means than the reso-nator techniques have been
used for a long time to study the broadband dynamic conductivity
of doped semiconductors. With a novel and advanced method of
measuring the broadband (100 MHz to 5 GHz) complex microwave
conductivity of that material class, we focus on the conductivity
power law in Si:P at $T$$\rightarrow$0 and at elevated
temperatures in a broad range of
donor concentration close to the MIT.\\

At concentrations of phosphorus in silicon below the critical
value of $n_c=3.5\times 10^{18}$ cm$^{-3}$, the donor electron
states are strongly localized due to disorder in Anderson sense
\cite{Anderson} and the corresponding wave functions resemble
those of a hydrogen atom \cite{ESbook}. Since some degree of
compensation by impurities of the opposite type is considered
inevitable, charge transport at low excitation energies is by
variable-range hopping between the donor sites, randomly
distributed in space \cite{ESbook,Mott}.\\

\subsection{Dynamic conductivity}

~~~The static conductivity\\ $\dc$ of the insulating Si:P vanishes
when $T \rightarrow 0$. The main issue we address is that of power
laws of the frequency-dependent conductivity in the variable-range
hopping regi- me at zero temperature:
\begin{eqnarray}
\quad \sig(\omega)\sim\omega^\alpha\,, \quad
\sigma=\sig+i\sigma_2\,.\label{power_law}
\end{eqnarray}
The theory of resonant photon absorption by pairs of states, one
of which is occupied by an electron and the other one is empty,
yields distinct limiting results for the conducti-
vity power law
in cases, where one of the relevant energy scales of the problem
dominates
over the others:\\

For the non-interacting system, where the photon ener-
gy prevails
over the electronic correlations, Mott has predicted a
sub-quadratic frequency-dependence of the conductivity
\cite{Mott}:
\begin{eqnarray}
\quad &&\sig(\omega)=\eta e^2 g^2 a \rhop^4 \quant^2\,,\label{Mott}\\
\quad &&\rhop=a\ln(2I_0/\quant)\nonumber\,.
\end{eqnarray}
Here, $\eta$ is a numerical coefficient \cite{ES,Mott}, $g$ is the
state density at the Fermi level, $a$ is the localization radius,
$\rhop$ is the most probable hopping distance and $I_0$ is the
pre-exponen-tial factor of the overlap integral between the
localized electron states.

Taking into account the Coulomb repulsion $\WW$ if both states in
a pair would be occupied by an electron,  Shklovskii and Efros
corrected the picture of the energy levels before and after a
photon is absorbed and derived $\sig(\omega)$ to be a sub-linear
function of frequency, as long as the Coulomb interaction term
dominates over the photon energy \cite{ES}:
\begin{eqnarray}
\quad &&\sig(\omega)=\eta e^2 g^2 a \rhop^4 \omega\,[\quant+\WW] \label{ES}\,,\\
\quad &&\WW=\frac{e^2}{\eps \rhop}\,.\nonumber
\end{eqnarray}
At higher frequencies, in the opposite limit, the sub-quadratic
behavior known from Mott (\ref{Mott}) is recovered. \\\\
The formula (\ref{ES}) is derived under the assumption of an
effectively constant density of states $g$ near the Fermi level.
This implies another characteristic energy of the system, the
Coulomb gap $\Cg$, to be small compared to $\WW$ and thus of no
significant effect on the states participating in the hopping
transport. The parabolic Coulomb gap forms in the density of the
impurity states around the Fermi level as another consequence of
the electronic correlations \cite{ESbook}. For the conductivity of
interacting electrons where the Coulomb term $\WW$ dominates over
the photon energy but falls inside the Coulomb gap, the reduction
of the density of states leads to a stronger, slightly
super-linear power law \cite{ES}:
\begin{eqnarray}
\quad \sig(\omega)\sim\omega/\ln(2I_0/\quant)
\label{inside_the_gap}\,.
\end{eqnarray}
\subsection{Dielectric function}

It is an additional advantage of a phase sensitive measurement to
gain the dielectric function $\eps$ from the imaginary part of the
complex conductivi-\\ty $\sigma=\sig+i\sigma_2$ \cite{Drs}:
\begin{eqnarray}
\quad
\varepsilon=\eps+i\varepsilon_2=1+i\frac{\sigma}{\varepsilon_0\omega}\,.
\label{epsilon}
\end{eqnarray}
We denote by $\varepsilon$ the full complex dielectric function of
Si:P,
 relative to the
permittivity $\varepsilon_0$ of vacuum, and use the SI units
throughout this text. As the MIT is approached upon doping $n$,
the localization radius $a$ diverges according to a power law
\cite{ESbook}:
\begin{eqnarray}
\quad a\sim|1-n/n_c|^{-\nu}\,. \nonumber
\end{eqnarray}
As a consequence, the electronic contribution to the dielectric
function is also expected to diverge following a power law when
the MIT is approached \cite{ESepsilon}:
\begin{eqnarray}
\quad \eps-\varepsilon_{\rm Si}\sim|1-n/n_c|^{-\zeta}\,,
\label{scaling}
\end{eqnarray}
where $\varepsilon_{\rm Si}=11.7$ is the dielectric constant of
the
host material Si.\\

\section{Experiments}

\subsection{Samples}
~~ Si:P-samples for this study were cut from a Czochralski-grown
cylindrical crystal \cite{crystal}, nominally uncompensated, with
a phosphorus concentration gradient along its axis. To remove
distorted surface layers \cite{distortion,Helgren}, the samples
were chemically and mechanically treated by well established
procedures. The donor concentration was determined from the
room-temperature resistivity \cite{Thurber} employing a commercial
four-probe measurement system (FPP 5000 by Veeco Instruments). For
high doping levels the resistivity ratio $\rho(4.2$K)/$\rho(300$K)
(determined from standard dc measurements) is consistent with
Ref.~\cite{Hornung}. In the present work, measurements on
Si:P-samples with relative donor concentration $n/n_c$ ranging
from 0.56 to 0.90 relative to the critical value at the MIT are
discussed, as summarized in Tab. \ref{n}.

\begin{table}[h]
  \caption{ Dopant concentration $n$ and the value of the
  room-temperature dc resistivity $\rho_{\rm dc}$ of the presented
  Si:P-samples. The second column lists the relative concentration
  with respect to the critical value $n_c=3.5\times 10^{18}$
  cm$^{-3}$ at the MIT. The power $\alpha$ of the conductivity power
  law $\sig(\omega)\sim\omega^\alpha$ and the full relative dielectric constant
  $\eps$ are also displayed in Fig.\ref{power} and Fig.\ref{epsilon}}\label{n}
  \begin{tabular}[htbp]{@{}ccccl@{}}
  &&\\
    \hline
    $n$  & $n/n_c$ & $\rho_{\rm dc}(300$ K) & $\alpha$ & $\eps$\\
($10^{18}$ cm$^{-3})$ &  &  ($\mathrm{\Omega}$ cm) &  &   \\
    \hline
    1.97  & 0.56  & 0.0162 & 1.13 & 23 \\
    2.29  & 0.65  & 0.0149 & 1.16 & 24.5 \\
    2.57  & 0.73  & 0.0139 & 1.10 & 28 \\
    2.91  & 0.83  & 0.0130 & 1.08 & 34 \\
    3.04  & 0.87  & 0.0127 & 1.05 & 41 \\
    3.14  & 0.90  & 0.0124 & 1.04 & 46 \\
    \hline
  \end{tabular}
\end{table}

\begin{figure}
  \includegraphics*[width=0.9\linewidth]{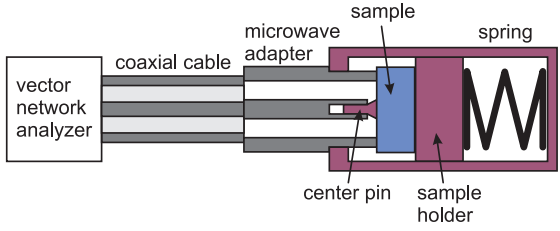}
  \caption{\small
Scematic of a broadband microwave spectrometer that employs a
vector network analyzer. A solid sample terminates the otherwise
open-ended coaxial line (after \cite{SD}).}\label{sceme}
\end{figure}

\newpage

\subsection{Broadband microwave spectroscopy}\label{technique}
~~ The broadband microwave spectrometer, previously employed for
measuring metallic samples \cite{SD,Anlage}, has been upgraded
\cite{RD} with microwave components, calibration technique and
evaluation procedure suitable to study the hopping transport in
doped semiconductors at temperature as low as possible.

The sample with a flat surface terminates the 1.2m-long coaxial
line, Fig.\ref{sceme}, needed to transport the microwave signal
from the source to the bottom of a $^4$He-pumped cryostat. The
reflected signal, containing all the useful information about the
sample, returns the same way back to the test set of a HP 8510
vector network analyzer. The measurement is phase sensitive and
yields the complex reflection coefficient. After an extensive
calibration procedure requiring three independent low temperature
measurements of calibration samples under reproduced conditions,
the reflection coefficient $S_{11}$ at the sample surface is
obtained. The complex impedance $Z$ is directly related to it via
\cite{Pozar}:
\begin{eqnarray}
\quad S_{11}=\frac{Z-Z_0}{Z+Z_0}\,\label{Z-S} \quad ,
\end{eqnarray}
with the characteristic impedance $Z_0$ of the microwave line.

In a next step, the complex conductivity $\sigma$ has to be
extracted from the impedance $Z$. If material is insulating, there
exists no direct solution, because the simple concepts known from
metallic samples \cite{Anlage} fail here. In the case of an
insulator, the surface impedance approach \cite{Drs}, useful for
thick metallic samples, yields a wrong (i.e. too strong) frequency
dependence of the conductivity $\sig$. If the semiconducting
sample is treated as a thin highly-conducting film, the
conductivity values are suspiciously large \cite{Lee}.

\subsection{Advanced data analysis}

\begin{figure}
\centering
  \includegraphics*[width=0.65\linewidth]{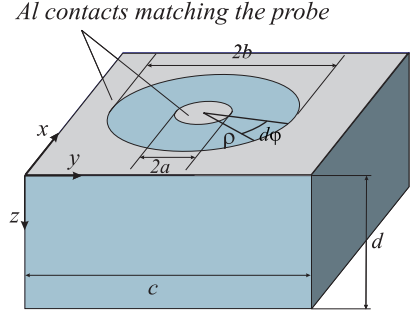}
  \caption{\small \; Sketch of the Si:P-sample with the aluminium contact-layer on top.
$2a=0.6$ mm, $2b=1.75$ mm, corresponding to a 2.4 mm microwave
adapter, $c=5$ mm, $d=2$ mm}\label{sample}

\end{figure}

For these reasons, we have developed a general and rigorous
solution for the problem. The electromagnetic wave penetrates deep
into a thick insulating sample, Fig.\ref{sample}, and forms a
three-dimensional distribution quite different from that of a
plane wave. Thus, we work with a solution for the field
distribution in an integral equation formulation, combined with
the variational principle \cite{LP}. Ascertaining, that the
electromagnetic field strength decreases below 1\% at the depth
corresponding to our sample thickness of 2 mm in the frequency
range evaluated, we forbear from taking into account the secondary
reflections at the back side of the sample as in the full-wave
analysis at higher frequencies considered by Brom and
collaborators \cite{Brom}. The resulting formula connects the
complex sample impedance $Z$ to the wave vector $\vec{k}$, the
latter containing the complex dielectric function $\varepsilon$ we
search for. (Note, that we hold to the convention
$\exp(i(\vec{k}\vec{x}-\omega t))$ for the Fourier-transformed of
the electromagnetic field in accordance with \cite{Drs,LP}. The
opposite sign convention is used in Ref. \cite{Brom,Misra} and by
the network analyzer.)
\begin{eqnarray}
&&\frac{Z_0}{Z}=\frac{-ik^2}{\pi
k_c\ln(b/a)}\int\limits_a^b\int\limits_a^b\int\limits_0^\pi\cos\varphi\,\frac{\exp(ik
r)}{r}\,\mbox{d}\varphi\, \mbox{d}\rho\, \mbox{d}\rho^{\,\prime}
,\nonumber\\
&&r=(\rho^2+\rho^{\,\prime2}-2\rho\rho^{\,\prime}\cos\varphi)^{1/2}\,
.\label{integral}
\end{eqnarray}

To solve this integral equation for $k$ we make use of the
practical ideas from \cite{Misra}. In a series expansion of the
exponential function the terms higher than the third term are
negligible below 5 GHz, integrating the rest one obtains a
quadratic equation in $\varepsilon=\eps+i\varepsilon_2$ \cite{RD}:
\begin{eqnarray}
\quad
\frac{1}{Z}=\frac{-i2\omega\varepsilon_0\varepsilon}{[\ln(b/a)]^2}\left[I_1-\frac{\omega^2\mu_0
I_3\varepsilon_0\varepsilon}{2}\right]\,   \label{qequation}
\end{eqnarray}
with the geometrical integrals $I_1=0.9084$ mm and $I_3=-0.2100$
mm$^3$ numerically evaluated for the dimensions $a$ and $b$ of our
coaxial probe as shown in Fig.\ref{sample}.\\

\subsection{Low-temperature ac measurements}

The frequency range from 100 MHz to 5 GHz for the reflection
coefficient measurements according to section \ref{technique}
contains 300 frequency points: 200 equidistant points from 0.1 GHz
to 1 GHz and 100 from 1 GHz to 5 GHz. The spectra are taken at the
base temperature 1.1 K of our $^4$He-pumped cryostat as well as at
elevated temperatures using a temperature controller Lake Shore
340 in an automatized procedure while heating the setup up to room
temperature. All the reported samples prove to be in the
zero-phonon regime at $T=1.1$~K because they show a saturation of
$\sig(T)$ and $\eps(T)$ in the whole frequency range as
$T\rightarrow 1$~K.

\begin{figure}
  \includegraphics*[width=\linewidth]{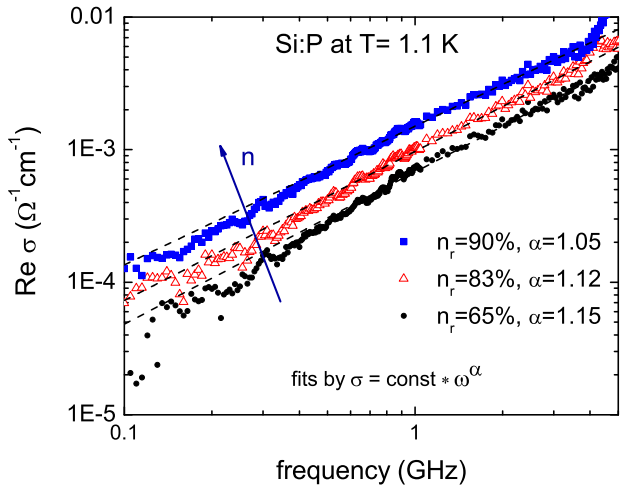} 
  \caption{ Typical spectra of the measured real part of the
  conductivity for Si:P with relative donor concentrations $n/n_c$ of 0.65, 0.83 and 0.90.
  The dashed lines are the fits by a two-parameter power law function.}\label{sigma_1K}
\end{figure}

\begin{figure}
  \includegraphics*[width=0.8\linewidth]{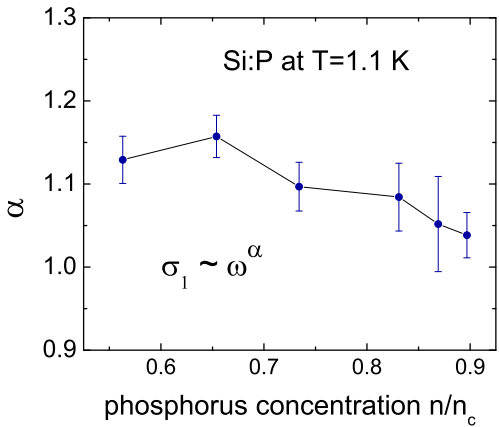} 
  \caption{Mean value of the conductivity power $\alpha$ obtained from the $\sig\sim\omega^{\alpha}$ fit. The dependence
  on the relative donor concentration $n/n_c$ becomes stronger as $n\rightarrow n_c=3.5\cdot 10^{18}$~cm$^{-3}$.}\label{power}
\end{figure}

\begin{figure}
  \includegraphics*[width=0.8\linewidth]{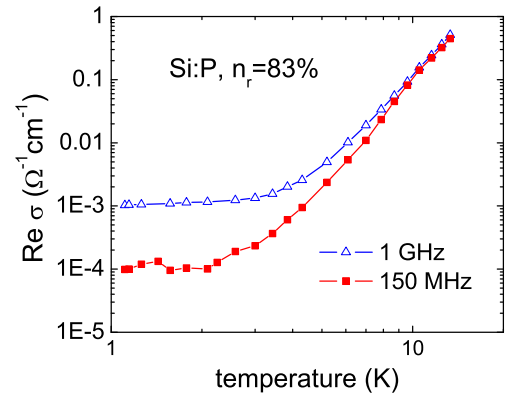} 
  \caption{Temperature dependence of the conductivity at fixed
  frequency on example of relative doping $n/n_c=$0.83, typical for
  all the investigated Si:P-samples.}\label{T_depend}
\end{figure}

\section{Results~and~discussion}

For each doping concentration  $n$ four to ten independent
measurements have been conducted. From the best spectra (lowest
noise and smallest influence of standing waves) mean values of the
conductivity power $\alpha$ and the dielectric constant $\eps$
have been determined.

\subsection{Frequency dependence of the conductivity in the zero-phonon
regime}

In Fig.\ref{sigma_1K} the measured real part of the
frequency-dependent conductivity is plotted on a log-log scale to
identify the power law. The fits by a two-parameter function
$\sig(\omega)=\mathrm{const}\cdot\omega^\alpha$ crossing the
origin are shown by the dashed lines. In Fig.\ref{power} the mean
values of the power $\alpha$ are plotted against the relative
dopant concentration. The frequency dependence of the conductivity
clearly follows a super-linear power law in the whole doping
range, where the exponent decreases slightly with doping; this
effect becomes even stronger when $n$ is increased further. From
this we infer, that hopping transport takes place deep inside the
Coulomb gap corresponding to equation (\ref{inside_the_gap}). A
super-linear conductivity power law was previously observed in
Si:As and Si:P by Castner and collaborators \cite{DC,MC} using
resonator techniques at certain frequency within the range of the
present work.  Our results are in accord with the measurements on
similar samples at higher frequencies (30~GHz to 3~THz) using
optical techniques \cite{Hering1,Hering2}. In contrast, a
sub-linear frequency dependence in the zero-phonon regime has been
reported by Lee and Stutzmann \cite{Lee} based on experiments on
Si:B in the microwave range and by Helgren et al. \cite{Helgren}
using quasi-optical experiments.

\subsection{Temperature dependence of the conductivity}

Leaving the zero-phonon regime by raising temperature, a gradual
increase of the conductivity $\sig$ is observed for all
investigated Si:P-samples. First, the temperature dependence is
approximately linear in agreement with the prediction of Austin
and Mott \cite{ES,AM}. With increasing $T$, it gradually becomes
stronger until the charge carrier activation into the conduction
band dominates over the hopping transport. The onset of the
thermal effects depends on the phosphorus concentration: the
higher the doping, the lower the temperature at which the
temperature dependence sets in. But the way it happens is the same
for all the samples measured. Taking the example of $n/n_c=0.83$,
we have plotted the conductivity values in Fig.\ref{T_depend} at
fixed frequencies of 150 MHz and 1 GHz. Above $T=9$~K the curves
merge into the thermally activated curve.

\begin{figure}
  \includegraphics*[width=\linewidth]{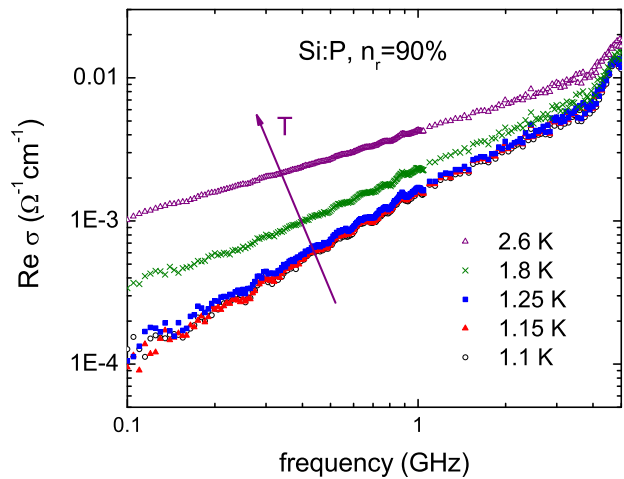}
  \caption{Temperature dependence of the conductivity spectra typical for
  all the investigated Si:P-samples.}\label{sigma_90}
\end{figure}

In fig.~\ref{sigma_90} the frequency dependent conductivity is
plotted for different temperatures, for the example of the crystal
with the high concentration $n/n_c=0.9$; the conductivity power
law gradually decreases with rising temperature. The transition to
a sub-linear power law upon raising $T$ is in accord with previous
observations \cite{MC,AM,PG}; to our knowledge, the theoretical
description of the gradual decrease of $\alpha$ is lacking.

\subsection{Dielectric function}

The dielectric function $\eps$ is independent of frequency in our
frequency range, taking the measurement uncertainty into account.
A fit with the function (\ref{scaling}) results in an exponent
$\zeta=$0.8, as shown in Fig.\ref{epsilon}.

\begin{figure}
  \includegraphics*[width=0.8\linewidth]{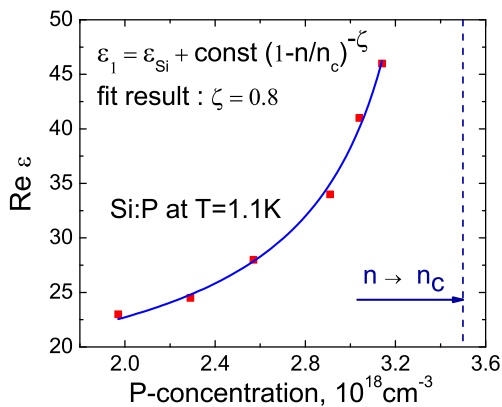} 
  \caption{Doping dependence of the low-temperature dielectric constant
  of Si:P. The solid line is the fit with
  the power law function (\ref{scaling}) for the electronic dielectric
  function $\eps-\varepsilon_{\rm Si}$.}\label{epsilon}
\end{figure}

\section{Conclusion}

Si:P-samples with relative doping $n/n_c$ from 56\% to 90\% show
temperature-independent dynamic response at the temperature of 1.1
K. Over a wide frequency range (100 MHz to 5 GHz) the conductivity
power law is slightly super-linear, with a faint decrease of the
power $\alpha$ upon doping. At higher doping, above 90~\%, the
conductivity power law drops abruptly. Interpretation and further
studies at lower base temperature using $^3$He are in progress. At
elevated temperatures the conductivity values are increasing and
the power $\alpha$ is falling gradually with rising $T$. The
electronic contribution to the dielectric function is constant and
increases with phosphorus concentration as expected when the MIT
is approached.

\begin{acknowledgement}
  We thank A.\,L. Efros, H.\,v. Löhneysen and K. Holczer for
  helpful discussions, A.\,W. Anajoh and M. Scheffler for the dc-measurements and die Landesgraduiertenför-\\derung
  Baden-Württemberg for the scolarship.
\end{acknowledgement}

\def\bstname{pss}

\end{document}